\documentclass[12pt]{article}
\usepackage{amsmath}
\usepackage{amsfonts}
\usepackage{makeidx}
\usepackage{amssymb}
\usepackage{times}
\usepackage{anysize}

\marginsize{3cm}{3cm}{2cm}{2cm}

\input{tcilatex}

\begin{document}

\title{Microscopic Foundations of Ohm and Joule's Laws -- The Relevance of
Thermodynamics}
\author{J.-B. Bru \and W. de Siqueira Pedra}
\maketitle

\begin{abstract}
We give a brief historical account on microscopic explanations of electrical
conduction. One aim of this short review is to show that Thermodynamics is
fundamental to the theoretical understanding of the phenomenon. We discuss
how the 2nd law, implemented in the scope of Quantum Statistical Mechanics,
can be naturally used to give mathematical sense to conductivity of very
general quantum many--body models. This is reminiscent of original ideas of
J.P. Joule. We start with Ohm and Joule's discoveries and proceed by
describing the Drude model of conductivity. The impact of Quantum Mechanics
and the Anderson model are also discussed. The exposition is closed with the
presentation of our approach to electrical conductivity based on the 2nd law
of Thermodynamics as passivity of systems at thermal equilibrium. It led to
new rigorous results on linear conductivity of interacting fermions. One
example is the existence of so--called AC--conductivity measures for such a
physical system. These measures are, moreover, Fourier transforms of time
correlations of current fluctuations in the system. I.e., the conductivity
satisfies, for a large class of quantum mechanical microscopic models,
Green--Kubo relations.
\end{abstract}


\section{Electrical Conductivity and Classical Physics}

\subsection{The Genesis of Ohm and Joule's laws}

G.S. Ohm was born in 1789 in Erlangen and came from a modest background (son
of a master locksmith). Nevertheless, he succeeded in learning basic
mathematics and sciences and became for about a decade teacher of
mathematics and physics in Cologne. During this time, he had been able to
elaborate his own experiments on electrical resistivity. He was originally
inspired by J. Fourier's work, published in 1822, about heat theory. Indeed,
G.S. Ohm drew a comparison of heat conduction with the electrical one in a
metallic bar. Based on this intuition, he published a few papers on
experimental outcomes about electrical resistivity in metals. He concluded
his work on electrical conduction by his famous theory \cite{thermo-ohm},
which was a theoretical deduction of his law from \textquotedblleft \emph{%
first principles}\textquotedblright . Indeed, he states that the current in
the steady regime is proportional to the voltage applied to the conducting
material. The proportionality coefficient is the conductivity (or inverse
resistivity) of the physical system. It is an empirical law which looks
almost obvious nowadays.

At that time, however, his writings were almost unknown. His book \cite%
{thermo-ohm} was at best completely ignored and at worst treated really
negatively. Rather than scientific, some critics were more ethical as they
were based on a priori conceptions on what science and nature are, probably
on what L. Daston and P. Galison have called \emph{truth--to--nature} \cite%
{objectivity}. Quoting \cite[p. 243]{history1}: \bigskip

\noindent ...\textit{Ohm's theory, to quote one critic, was
\textquotedblleft a web of naked fancies\textquotedblright , which could
never find the semblance of support from even the most superficial
observation of facts; \textquotedblleft he who looks on the
world\textquotedblright , proceeds the writer, \textquotedblleft with the
eye of reverence must turn aside from this book as the result of an
incurable delusion, whose sole effort is to detract from the dignity of
nature\textquotedblright . ... where he had looked for commendation he found
at best complete indifference, and at worst open abuse and derision. ... The
influence of this opposition} (some school official)\textit{\ reached the
Minister of Education himself, and he, speaking officially, definitely
pronounced it as his opinion that \textquotedblleft a physicist who
professed such heresies was unworthy to teach science\textquotedblright .}%
\bigskip

Retrospectively, such comments in \textquotedblleft a country so well
represented in the world of science by men of eminence and
knowledge\textquotedblright  ~\cite{history1} are marks of revolutionary
ideas, but it was a real bitter blow for G.S. Ohm: He gave up his teacher
position at Cologne and started six years of hard times. His work was
nevertheless occasionally cited and rumors about Ohm's theory started to
appear in different places. This includes America where the famous physicist
J. Henry asked in 1833 his colleagues: \textquotedblleft Could you give me
any information about the theory of Ohm? Where is it to be
found?\textquotedblright\ J. Henry succeeded in having this information by
going to England in 1837 at a time when Ohm's work had already become
famous, particularly outside his own country.

Although at the origin of Ohm's intuition, the relation between heat and
electrical conduction has not been established by himself, but J.P. Joule,
who was born in 1818 in England. The pivotal ingredient was the wide concept
of energy. Joule's intuition seems to have been that the different physical
properties appearing in nature can be tracked by the concept of energy. He
thus studied different forms of energy in order to relate them. The
conversion of mechanical work into heat is a famous topic of such studies.
His works, although also very controversial at the beginning, were seminal
and yielded the \emph{1st law of Thermodynamics}, see, e.g.,
~\cite{Thermo1}%
. Recall also that all mechanical work can be converted to heat but the
converse is not true, in general. This observation refers to the \emph{2nd
law of Thermodynamics} and the concept of \emph{entropy} invented by R.J.E.
Clausius in 1865.

Applied to electricity theory, Joule's intuition allowed to establish a
relation between heat and electrical conduction. Indeed, more than one
decade after Ohm's discovery \cite{thermo-ohm} on linear electrical
conduction, the physicist J. P. Joule observed \cite{J} in 1840 that the heat
(per second) produced within an electrical circuit is proportional to the
electrical resistance and the square of the current:\bigskip

\noindent ...\textit{the calorific effects of equal quantities of
transmitted electricity are proportional to the resistances opposed to its
passage, whatever may be the length, thickness, shape, or kind of metal
which closes the circuit: and also that, coeteris paribus, these effects are
in the duplicate ratio of the quantities of transmitted electricity; and
consequently also in the duplicate ratio of the velocity of transmission. }%
\smallskip

\hfill \lbrack Joule, 1840]\bigskip

Nowadays, electrical conductivity usually refers to Ohm and Joule's laws.
They are indeed among the most resilient laws of (classical) electricity
theory. Materials are called ohmic or nonohmic, depending on whether they
obey Ohm's law. Both assertions are empirical laws and, as usual, they
generated at least as many theoretical problems as they solved. From a
mathematically rigorous point of view, the microscopic origin of the
phenomenological behavior of macroscopic conductors described by these laws
is still not completely understood, specially in the DC regime. Moreover, as
recent experiments show, Ohm's law is not only valid at macroscopic\emph{\ }%
scales. Indeed, the validity of Ohm's law at the atomic scale for a purely
quantum system has experimentally been verified \cite{Ohm-exp} in 2012. Such
a behavior was unexpected \cite{Ohm-exp2}:\bigskip

\noindent \textit{...In the 1920s and 1930s, it was expected that classical
behavior would operate at macroscopic scales but would break down at the
microscopic scale, where it would be replaced by the new quantum mechanics.
The pointlike electron motion of the classical world would be replaced by
the spread out quantum waves. These quantum waves would lead to very
different behavior. ... Ohm's law remains valid, even at very low
temperatures, a surprising result that reveals classical behavior in the
quantum regime. }\smallskip

\hfill \lbrack D.K. Ferry, 2012]

\subsection{Towards a microscopic theory of electrical conduction}

In the end of the nineteenth century, the so--called classical physics
reached a very high level of completeness with Classical Mechanics,
Electrodynamics, and Thermodynamics. However, borderline problems became
manifestly more and more important and eventually led to the scientific
revolution of the beginning of the twentieth century. For instance, the
study of the link between Classical Mechanics and Thermodynamics yielded the
so--called Statistical Physics via Gibbs and Boltzmann's genius intuitions.

Classical Mechanics is indeed a causal theory based on elementary physical
objects satisfying Newton's laws of motion. The exceptional success of this
theory together with new technologies like photography had propagated a new
view point on science during the last part of the nineteenth century with
the so--called \emph{mechanical objectivity} \cite{objectivity}. By contrast,
Thermodynamics emphasizes the concepts of energy and entropy of macroscopic
systems. It speaks about reversible and irreversible processes, but it does
not care about the concrete system under consideration. Classical Mechanics
is in some sense a bottom--up or \textquotedblleft local\textquotedblright\
approach, whereas Thermodynamics is a top--down or global one.

In order to bridge the gap between both theories, L. Boltzmann successfully
tried to go from Classical Mechanics towards Thermodynamics via statistical
or probabilist methods. For more details on Boltzmann's legacy, see
~\cite{Boltzman}%
. His $H$--theorem (1872) was undoubtedly an important achievement \cite%
{Boltzmanbisbis} as it has provided a mechanical explanation of the 2nd law
of Thermodynamics from the dynamics of rarefied gases. Boltzmann's vision of
\textquotedblleft atoms\textquotedblright\ as the physical objects
satisfying Newton's laws was however again very controversial for a long time%
\cite{Boltzmanbis}: \textquotedblleft have you seen any?\textquotedblright\
might have said the famous physicist and philosopher E. Mach as a reply to
the issue of atoms (cf. \cite[p. 20]{Boltzmanbis}). E. Mach had indeed a
philosophical approach centered on the world of sensations in a similar
spirit of mechanical objectivity, whereas L. Boltzmann also focussed on
mathematical structures. See later the development of \emph{structural
objectivity}\cite{objectivity} (M. Planck (1906), B. Russell, H. Poincar\'{e}%
, C.S. Peirce, etc.). Similar ethical oppositions appeared in other
sciences: S. Ram\'{o}n y Cajal and C. Goldi were together Nobel laureates in
1906, but C. Goldi violently opposed S. Ram\'{o}n y Cajal's theory of
neurons (similar to Boltzmann's theory of atoms) to explain the global
system which is the brain. For more details, see
~\cite{cajal-goldi}%
. As explained in
~\cite{objectivity}%
, the opposite conceptions of science were in this case truth--to--nature
(Goldi) and mechanical objectivity (Ram\'{o}n y Cajal) as well as continuous
(Goldi) versus discontinuous (Ram\'{o}n y Cajal) visions.

In the same spirit as Boltzmann, it was natural to rise the question of the
microscopic origin of Ohm and Joule's laws. In 1846, W. Weber conjectured
that currents were a flow of charged fluids and in 1881, H. von Helmholtz
argued the existence of positive and negative charges as \textquotedblleft
atoms of electricity\textquotedblright . The discovery of the electron took
place in the last years of the nineteenth century by J.J. Thomson (Nobel
Prize in Physics 1906) and others. It is the first discovered elementary
particle. Based on the vision that current is a flow of electrons, the
celebrated Drude model was next proposed \cite{drude} in 1900 to give a
mechanical explanation of the phenomenon of conductivity. This model or its
extension, the Drude--Lorentz model (1905), are still used as microscopic
explanations of conductivity in textbooks. Indeed, although the motion of
electrons and ions is treated classically and the interaction between these
two species is modeled by perfectly elastic random collisions, this quite
elementary model provides a qualitatively good description of DC-- and
AC--conductivities in metals.

\section{Electrical Conductivity and Quantum Mechanics}

\subsection{Emergence of Quantum Mechanics}

The main principles of physics were considered as well--founded by the end
of the nineteenth century, even with, for instance, no satisfactory
explanation of the phenomenon of thermal radiation, first discovered in 1860
by G. Kirchhoff. In contrast to classical physics, which deals with
continuous quantities, Planck's intuition was to introduce an intrinsic
discontinuity of energy and a unsual \footnote{in regards to Boltzmann's studies, which meanwhile have strongly influenced Planck's work. In modern terms Planck used the celebrated Bose--Einstein statistics.}
 statistics (without any conceptual foundation, in a ad hoc way) to explain thermal radiation. Assuming the existence
of a quantum of action $h$, the celebrated Planck's constant, and this pivotal statistics he derived the
well--known Planck's law of thermal radiation. Inspired by Planck's ideas,
Einstein presented his famous discrete (corpuscular) theory of light to
explain the photoelectric effect.

Emission spectra of chemical elements had also been known since the
nineteenth century and no theoretical explanation was available at that
time. It became clear that electrons play a key role in this phenomenon.
However, the classical solar system model of the atom failed to explain the
emitted or absorbed radiation. Following again Planck's ideas, N. Bohr
proposed in 1913 an atomic model based on discrete energies that
characterize electron orbits. It became clear that the main principles of
classical physics are unable to describe atomic physics.

Planck's quantum of action, Einstein's quanta of light (photons), and Bohr's
atomic model could not be a simple extension of classical physics, which, in
turn, could also not be questioned in its field of validity. N. Bohr tried
during almost a decade to conciliate the paradoxical looking microscopic
phenomena by defining a radically different kind of logic. Bohr's concept of
complementarity gave in 1928 a conceptual solution to that problem and
revolutionized the usual vision of nature. See, e.g.,
~\cite{chevalley}%
. Classical logic should be replaced by quantum logic as claimed \cite%
{BvonNeu} by G. Birkhoff and J. von Neumann in 1936.

On the level of theoretical physics, until 1925, quantum corrections were
systematically included, in a rather \emph{ad hoc} manner, into classical
theories to allow explicit discontinuous properties. Then, as explained for
instance in
~\cite{shrodinger}%
, two apparently complementary directions were taken by W.K. Heisenberg and
E. Shr\"{o}dinger, respectively, to establish basic principles of the new
quantum physics, in contrast with the \textquotedblleft old quantum
theory\textquotedblright\ starting in 1900. Indeed, even with the so--called
correspondence principle of N. Bohr, \textquotedblleft many problems, even
quite central ones like the spectrum of helium atom, proved inaccessible to
any solution, no matter how elaborate the conversion\textquotedblright , see
\cite[p. 18]{shrodinger}.

\subsection{Quantum Fermi liquids}

Electric current is carried by electrons, purely quantum objects (W.E.
Pauli, 1925; E. Fermi, 1925; P.A.M. Dirac, 1929), whereas the Drude model
describes \emph{non--interacting classical} particles interacting with
impurities via perfectly elastic collisions. Quantum Mechanics, which
governs the microscopic world, represents a radical transformation of usual
principles of classical physics and it is therefore not at all satisfactory
to see the Drude (or the Drude--Lorentz) model as a proper microscopic
explanation of conductivity, even with good agreements with experimental
data. As one can see from the existence of superconducting phases first
discovered in 1911, electrons can show collective behaviors while satisfying
the celebrated Pauli exclusion principle.

A. Sommerfeld and H. Bethe modified in 1933 the Drude model to take into
account quantum effects. Essentially, they replaced the classical
point--like particles of Drude, carriers of electrical current, with
fermions. In particular, the carriers present quantum coherences and obey
the Fermi--Dirac statistics. However, the Drude--Sommerfeld model describes
a system of non--interacting fermions although electrons strongly interact
with each other via the Coulomb repulsion. A formal explanation of the
success of this model is given \cite{Landau} by L.D. Landau in the fifties.
His theory is based on the concept of \emph{Landau Fermi liquids} (or Fermi
liquids).

Landau's idea is, in a caricatured view, that the low--energy exited states
of a Fermi system with interparticle interactions can be mapped onto the
states of an effective non--interacting (or ideal) Fermi system. The
theoretical justification of such a behavior, i.e., the fact that the
electron--electron scattering remains negligible to change the momentum
distribution, results from the Pauli exclusion principle for energies near
the Fermi level. More precisely, if the system is at initial time in a state
closed to an ideal system (weakly excited), then its time--dependent state
can be uniquely described by occupation numbers of \emph{quasiparticles} (as
approximated quantum numbers). Moreover, L.D. Landau postulates the
existence of a function $\mathrm{f}_{k,k^{\prime }}$, the so--called Landau
interaction function, which quantifies the energy change of a quasiparticle
of quasimomentum $k$ in the presence of a quasiparticle of quasimomentum $%
k^{\prime }$. The effective mass, another parameter of Fermi liquids,
determines the dispersion relation of quasiparticles, i.e., the energy of
quasiparticles as a function of their quasimomenta. This effective (or
phenomenological) theory has been very successful in explaining the physics
of some electron systems, called Fermi liquids. Fermion systems are called
non--Fermi liquids if their behavior does not correspond to Landau's
predictions. Non--Fermi liquid behaviors usually appear in low dimensions.
For instance, in one dimension, the celebrated Luttinger liquid replaces the
(Landau) Fermi liquid. For more details, see
~\cite{dia-current}%
.

\subsection{From theoretical physics to mathematics: The Anderson model}

Resistivity of metals is believed to be due to interparticle interactions
but also to inhomogeneities of the conducting crystal. Disordered electron
liquids are therefore an important issue in this context. The theory of
Fermi liquids can be extended to disordered systems but major differences
appear as compared to the (space) homogeneous systems. New properties like
the so--called \emph{Anderson localization} are consequence of strong space
inhomogeneities, even in the absence of interparticle interactions.

Anderson localization corresponds to the absence of electron transport at
strong disorder and has been predicted \cite{Anderson} by the physicist P.W.
Anderson in 1958. This allows to guess a metal--insulator transition in
three dimensions. This theory has experimentally been investigated and P.W.
Anderson, together with N.F. Mott and J.H. van Vleck, won the 1977 Nobel
price in physics for \textquotedblleft their fundamental theoretical
investigations of the electronic structure of magnetic and disordered
systems\textquotedblright .

The Anderson model corresponds to a single quantum particle within a random
potential. It is one of the most important types of random Schr\"{o}dinger
operators, which constitute nowadays an advanced and relatively mature
branch of mathematics. In fact, random Schr\"{o}dinger operators start to be
studied in the seventies. The Anderson localization for an one--dimensional
model was first proved by I. Goldsheid, S. Molchanov and L. Pastur in 1977,
while a similar result for the one--dimensional Anderson model was obtained
in 1981 by H. Kunz and B. Souillard. It is known that, in general, the
one--dimensional Anderson model only has purely point spectrum with a
complete set of localized eigenstates (Anderson localization) and it is thus
believed that no steady current can exist in this case. For more detailed
introduction to the Anderson model and more general random Schr\"{o}dinger
operators, see for instance the lecture notes of
~\cite{WernerKirsch}%
.

Nevertheless, mathematical studies usually focus on the existence of
(dynamical or spectral) Anderson localizations and, even in absence of
interactions, there are only few mathematical results on transport
properties of such random models that yield Ohm's law in some form.

In 2007, A. Klein, O. Lenoble and P. M\"{u}ller introduced \cite{Annale} for
the first time the concept of a \textquotedblleft conduc%
\-%
tivity measure\textquotedblright\ for a system of non--interacting fermions
subjected to a random potential. More precisely, the authors considered the
Anderson tight--binding model in presence of a time--dependent spatially
homogeneous electric field that is adiabatically switched on. The fermionic
nature of charge carriers -- electrons or holes in crystals -- as well as
thermodynamics of such systems were implemented by choosing the Fermi--Dirac
distribution as the initial density matrix of particles. In
~\cite{Annale}
only systems at zero temperature with Fermi energy lying in the localization
regime are considered, but it is shown in
~\cite{JMP-autre}
that a conductivity measure can also be defined without the localization
assumption and at any positive temperature. Their study can thus be seen as
a mathematical derivation of Ohm's law for space--homogeneous electric
fields having a specific time behavior.

~\cite{Cornean}
is another mathematical result on free fermions proving Ohm's law for
graphene--like materials subjected to space--homogeneous and time--periodic
electric fields. Joule's law and heat production are not considered, at
least not explicitly, in these mathematical studies.

\section{Electrical Conductivity and 2nd Law of Thermodynamics}

Via
~\cite{Annale,JMP-autre}
one sees that measures (instead of functions or other types of
distributions) are the natural mathematical objects to be used to describe
conductivity starting form microscopic quantum dynamics. We claim that this
is so because of the 2nd law of Thermodynamics. Indeed, such a principle
guarantees the positivity of certain quadratic forms on external electric
fields, which naturally appears by considering linear response. By Bochner's
theorem, in a convenient form, such quadratic forms define measures. In the
case of current response to external electric fields, one gets
AC--conductivity measures. This approach permits to tackle the mathematical
problem of a rigorous microscopic description of the phenomenon of linear
conductivity starting from first principles. Moreover, it is general enough
to be applied to interacting systems. We implement the 2nd law of
Thermodynamics in the scope of algebraic Quantum Mechanics, by using the
remarkable results \cite{PW} of W. Pusz and S. L. Woronowicz: We consider
the 2nd law as a first principle of Physics which supplements Quantum
Mechanics in the sense that it singles out special states of the considered
systems. Indeed, states of infinite systems that are compatible with the 2nd
law exist for a huge class of dynamics.

In fact, the 2nd law is \textquotedblleft \textit{one of the most perfect
laws in physics}\textquotedblright\ \cite[Section 1]{lieb-yngvasonPhysReport}
and it has never been faulted by reproducible experiments. Its history
starts with works of S. Carnot in 1824. Different popular formulations of
the same principle have been stated by R.J.E. Clausius, W. Thomson or Lord
Kelvin (and M. Planck), and C. Cara%
\-%
th\'{e}odory. Our study is based on Kelvin--Planck statement while avoiding
the concept of \textquotedblleft cooling\textquotedblright\ \cite[p. 49]%
{lieb-yngvasonPhysReport}: \bigskip

\noindent \textit{No process is possible, the sole result of which is a
change in the energy of a simple system (without changing the work
coordinates) and the raising of a weight. }\bigskip

Using this formulation of the 2nd law, we define the concept of \emph{%
thermal equilibrium} states by using algebraic Quantum Mechanics as
mathematical framework. It is a well--known approach -- originally due to J.
von Neumann (cf. von Neumann algebras, $C^{\ast }$--algebras) -- that
extends the Hilbert space formulation of Quantum Mechanics. One important
result of the theory of $C^{\ast }$--algebras, obtained in the forties, is
the celebrated GNS (Gel'fand--Naimark--Segal) representation of states,
which permits a natural relation between the new algebraic formulation and
the usual Hilbert space based formulation of Quantum Mechanics to be
established. Indeed, I.E. Segal proposed to leave the Hilbert space approach
to consider quantum observables as elements of certain involutive Banach
algebras, now known as $C^{\ast }$--algebra. The GNS representation has also
led to very important applications of the Tomita--Takesaki theory, developed
in seventies, to Quantum Field Theory and Statistical Mechanics. These
developments mark the beginning of the algebraic approach to Quantum
Mechanics and Quantum Field Theory. For more details, see, e.g.,
~\cite{Emch}%
.

The algebraic formulation turned out to be extremely important and fruitful
for the mathematical foundations of Quantum Statistical Mechanics and have
been an important branch of research during decades with lots of works on
quantum spin and Fermi systems. See, e.g.,
~\cite{BratteliRobinson,Israel}
(spin) and
~\cite{Araki-Moriya,BruPedra2,BruPedra-homog}
(Fermi). Basically, it uses some $C^{\ast }$--algebra $\mathcal{X}$, the
self--adjoint elements of which are the so--called observables of the
physical system. States on the $C^{\ast }$--algebra $\mathcal{X}$ are, by
definition, continuous linear functionals $\rho \in \mathcal{X}^{\ast }$
which are normalized and positive, i.e., $\rho (\mathbf{1})=1$ and $\rho
(A^{\ast }A)\geq 0$ for all $A\in \mathcal{X}$. They represent the state of
the physical system.

To conveniently define equilibrium states in our case,
~\cite{PW}
is pivotal because it gives a definition of equilibrium by using the
Kelvin--Planck statement via the notion of \emph{passive} states: The
internal dynamics of the system is a strongly continuous one--parameter
group $\tau \equiv \{\tau _{t}\}_{t\in {\mathbb{R}}}$ of $\ast $%
--automorphisms of $\mathcal{X}$ with (generally unbounded) generator $%
\delta $. Usually, $\delta $ is a dissipative and closed derivation of $%
\mathcal{X}$. On this system, one applies a \emph{cyclic} process of length $%
T\geq 0$, that is, a continuously differentiable family $\{A_{t}\}_{t\in
\mathbb{R}}\subset \mathcal{X}$ of self--adjoint elements of $\mathcal{X}$
such that $A_{t}=0$ for all $t\leq s$ and $t\geq T+s$. The perturbed
dynamics is the solution $\{\tau _{t,s}\}_{t\geq s}$ of the non--autonomous
evolution equation defined, for any $B\in \mathrm{Dom}(\delta )$, by
\[
\forall s,t\in {\mathbb{R}},\ t\geq s:\quad \partial _{t}\tau _{t,s}\left(
B\right) =\tau _{t,s}\left( \delta \left( B\right) +i\left[ A_{t},B\right]
\right) ,\quad \tau _{s,s}\left( B\right) :=B\ .
\]%
The state of the system evolves as $\rho _{t}=\rho \circ \tau _{t,s}$ for
any $t\geq s$ at fixed initial state $\rho \in \mathcal{X}^{\ast }$. Then,
as explained in \cite[p. 276]{PW}, a state $\rho \in \mathcal{X}^{\ast }$ is
\emph{passive} iff the full work performed by the external device is
non--negative for all cyclic process $\{A_{t}\}_{t\geq s}\subset \mathcal{X}$
of any time--length $T\geq 0$, i.e.,
\begin{equation}
L_{\rho }^{A}:=\int_{s}^{T}\rho \circ \tau _{t,s}\left( \partial
_{t}A_{t}\right) \mathrm{d}t\geq 0\ .  \label{work}
\end{equation}%
In this way the Kelvin--Planck statement of the 2nd law can be formulated in
precise mathematical terms. If the product state $\otimes _{j=1}^{n}\rho $
is passive for any $n\in \mathbb{N}$ copies $(\mathcal{X}_{1},\tau _{1},\rho
_{1}),$ $\ldots ,(\mathcal{X}_{n},\tau _{n},\rho _{n})$ of the original system
defined by $(\mathcal{X},\tau ,\rho )$, then $\rho $ is called \emph{%
completly passive} \cite{PW}. Such states are the \emph{thermal equilibrium
states} of our setting. \cite[Theorem 1.4]{PW} shows that thermal
equilibrium states in this sense are exactly the \emph{KMS} \cite%
{BratteliRobinson} (Kubo--Martin--Schwinger) states of the corresponding $%
C^{\ast }$--dynamical system.

In our approach to electrical conduction, the $C^{\ast }$--algebra $\mathcal{%
X}$\ is the CAR algebra associated to the $d$--dimensional cubic lattice $%
\mathfrak{L}:=\mathbb{Z}^{d}$ ($d\in \mathbb{N}$) and particles of finite
spin. The initial state is a thermal equilibrium state and cyclic processes
are induced by electromagnetic potentials $\{\eta A_{t}\}_{t\geq s}$ with
constant strength $\eta \geq 0$ within some finite region $\Lambda $ of the
lattice $\mathfrak{L}$. This yields to perturbed dynamics with discrete
magnetic Laplacians in disordered media (like in the Anderson model). The
quadratic response with respect to $\eta $ of the full heat production or
electromagnetic work per unit volume turns out to equal $Q_{\rho
}^{A}=\varphi (A\ast \tilde{A})$, where $\varphi $ is a distribution, $%
\tilde{A}(t):=A(-t)$ and $A\in C_{0}^{\infty }(\mathbb{R},\mathbb{R})$.
Indeed, we have shown \cite{OhmII, OhmV} that $|\Lambda |^{-1}L_{\rho }^{\eta
A}-\eta ^{2}Q_{\rho }^{A}$ is of order $\mathcal{O}(\eta ^{3})$, uniformly
w.r.t. $A$ and the size $|\Lambda |$ of the region $\Lambda $ where the
electromagnetic field is applied. The 2nd law, that is, (\ref{work}),
implies then that $Q_{\rho }^{A}=\varphi (A\ast \tilde{A})\geq 0$, i.e., $%
\varphi $ is a \emph{distribution of positive type}. By the Bochner--Schwarz
theorem, there is a positive measure $\tilde{\mu}$ on $\mathbb{R}$ such that
\[
Q_{\rho }^{A}=\int_{\mathbb{R}}\mathrm{d}\tilde{\mu}(\nu )|\hat{A}(\nu
)|^{2}=\int_{\mathbb{R}\backslash \{0\}}\mathrm{d}\tilde{\mu}(\nu )\nu ^{-2}|%
\hat{E}(\nu )|^{2}
\]%
for all $A\in C_{0}^{\infty }(\mathbb{R},\mathbb{R})$. Here, $E=-\partial
_{t}A$ is the electric field in the Weyl gauge and $\hat{A},\hat{E}$ are the
Fourier transforms of $A,E$ with support outside some neighborhood of $\nu =0
$.

The measure $\mathrm{d}{\mu }(\nu ):=\nu ^{-2}\mathrm{d}\tilde{\mu}(\nu )$
on $\mathbb{R}\backslash \{0\}$ turns out to be the AC--conductivity measure
we are looking for and the quantity $\mathrm{d}\mu (\nu )|\hat{E}(\nu )|^{2}$
is the heat production due to the component $\hat{E}(\nu )$ of frequency $%
\nu $ of the electric field $E$, in accordance with Joule's law. It is
directly related to the passivity property of thermal equilibrium states on
the CAR algebra. In other words, the existence of AC--conductivity measures
results from the 2nd law of Thermodynamics in connection with the full
quantum microscopic dynamics of the considered system.\emph{\ }Moreover, the
approach to linear (or quadratic in the energetic view point) response we
propose has also the following technical and conceptual advantages, even in
the non--interacting case:

\begin{itemize}
\item The conductivity measure naturally appears as the Fourier transform of
current--current time correlations, that is,\ four--point correlation
functions, in this framework. This means that Green--Kubo relations are
generally valid, from first principles.\emph{\smallskip }

\item The algebraic formulation allows a clear link between macroscopic
transport properties of fermion systems and the CCR algebra of current
fluctuations associated to that systems. The latter is related to
non--commutative central limit theorems.\emph{\smallskip }

\item Moreover, this approach can be naturally used to define and analyze
conductivity measures for \emph{interacting} fermions as well.\emph{%
\smallskip }
\end{itemize}

In
~\cite{OhmI,OhmII}
we study free lattice fermions subjected to a static bounded potential and a
time-- and space--dependent electromagnetic field.
~\cite{OhmI,OhmII}
establish a form of Ohm and Joule's laws valid at microscopic scales,
uniformly with respect to the size of the region on which the electric field
is applied. It is in accordance with the validity of Ohm's law in the
quantum world, i.e., at microscopic scales, see
~\cite{Ohm-exp}%
.
~\cite{OhmIII,OhmIV}
extend the results of
~\cite{OhmI,OhmII}
to macroscopic scales and are reminiscent of
~\cite{Annale,JMP-autre}%
. For more details, see the discussions in
~\cite{OhmIII}%
. Part of the phenomenology of the Drude model can be derived from our more
detailed study \cite{OhmIV} of macroscopic conductivity measures of
free--fermions in disordered media.

Therefore,
~\cite{OhmI,OhmII,OhmIII,OhmIV}
give a complete, mathematically rigorous, microscopic derivation of the
phenomenon of linear conductivity from first principles of Thermodynamics
and Quantum Mechanics, only. These studies are restricted to
non--interacting fermion systems. However, it is believed in theoretical
physics that electric resistance of conductors should also result from the
interactions between charge carriers, and not only from the presence of
inhomogeneities (impurities). In
~\cite{OhmVbis,OhmV,OhmVI}%
, we succeed to extend our previous results to fermion systems with
interactions.

To conclude, we think that such an approach can be useful in other contexts
since it gives appropriate tools to tackle mathematically what is known in
Physics as \emph{excitation spectrum}. Indeed, the concept of excitation
spectrum is usually associated with the spectrum of a self-adjoint operator
describing the energy of the system. In condensed mater physics, this notion
mainly comes from superfluid helium 4, a quantum boson liquid which can be
described by the spectrum of collective excitations via the celebrated
Bogoliubov theory \cite{BZ}. However, there is a plethora \cite%
{spectrumexcitation1,spectrumexcitation2} of other types of elementary
excitations not covered by the Bogoliubov theory. We show \cite%
{OhmII,OhmIII,OhmV} that the notion of conductivity measure we defined is
nothing else than a spectral measure of the generator of dynamics in an
appropriate representation.


\begin{thebibliography}{99}
\bibitem{thermo-ohm} G.S. Ohm, \emph{Die galvanische Kette, mathematisch
bearbeitet} (Riemann, Berlin, 1827).

\bibitem{objectivity} L.J. Daston and P. L. Galison, \emph{Objectivity}
(Zone Books, 2007).

\bibitem{history1} I.B. Hart, \emph{Makers of Science} (Oxford University
Press, London, 1923).

\bibitem{Thermo1} G. Emch, C. Liu, \emph{The Logic of Thermostatistical
Physics} (Springer-Verlag, New York, 2002).

\bibitem{J} J. P. Joule, \emph{Philosophical Transactions of the Royal
Society of London} \textbf{4}, 280 (1840).

\bibitem{Ohm-exp} B. Weber et al., \emph{Science} \textbf{335}(6064), 64
(2012).

\bibitem{Ohm-exp2} D.K. Ferry, \emph{Science} \textbf{335}(6064), 45 (2012).

\bibitem{Boltzman} G. Gallavotti et al., \emph{Boltzmann's Legacy} (ESI
Lectures in Math. \& Phys., EMS, 2007)

\bibitem{Boltzmanbisbis} C. Villani, \emph{H-theorem and beyond: Boltzmann's
entropy in today's mathematics}. In ESI Lectures in Math. \& Phys., EMS, Ed.
G. Gallavotti et al., 2007, p. 129--144.

\bibitem{Boltzmanbis} J. Renn, \emph{Boltzmann and the end of the
mechanistic worldview}. In ESI Lectures in Math. \& Phys., EMS, Ed. G.
Gallavotti et al., 2007, p. 7-26.

\bibitem{cajal-goldi} J.A. De Carlos and J. Borrell, \emph{Brain Research
Reviews} \textbf{55}, 8 (2007).

\bibitem{drude} P. Drude, Annalen der Physik \textbf{306} (3), 566 (1900);
ibid. \textbf{308}(11), 369 (1900).

\bibitem{chevalley} N. Bohr, \emph{Physique atomique et connaissance humaine}%
, translation by E. Bauer and R. Omn\`{e}s, Ed. C. Chevalley (Paris,
Editions Gallimard, 1991).

\bibitem{BvonNeu} G. Birkhoff and J. Von Neumann, \emph{Annals of Mathematics%
} \textbf{37}(4), 823 (1936).

\bibitem{shrodinger} J. Renn, \emph{Schr\"{o}dinger and the genesis of wave
mechanics}. In ESI Lectures in Math. \& Phys., EMS, Ed. W.L. Reiter, J.
Yngvason, 2013, p 9--36.

\bibitem{Landau} L.D. Landau, \emph{Zh. Eksp. Teor. Fiz.} \textbf{30}, 1058
(1956) [\emph{Sov. Phys. JETP} \textbf{3}, 920 (1957)]; \emph{Zh. Eksp.
Teor. Fiz.} \textbf{32}, 59 (1957) [\emph{Sov. Phys. JETP} \textbf{5}, 101
(1957)]

\bibitem{dia-current} G.F. Giuliani and G. Vignale, \emph{Quantum Theory of
the Electron Liquid} (Cambrigde Univ. Press., 2005).

\bibitem{Anderson} P.W. Anderson, \emph{Phys. Rev.} \textbf{109}(5), 1492
(1958).

\bibitem{WernerKirsch} Werner Kirsch, \emph{An invitation to random Schr\"{o}%
dinger operators} (with an appendix by F. Klopp). In Panoramas et Synth\`{e}%
ses, Soci\'{e}t\'{e} Math\'{e}matique de France \textbf{25}, 2008.

\bibitem{Annale} A. Klein, O. Lenoble, and P. M\"{u}ller, \emph{Annals of
Mathematics} \textbf{166}, 549 (2007).

\bibitem{JMP-autre} A. Klein and P. M\"{u}ller, \emph{J. Math. Phys.,
Analysis, Geometry} \textbf{4}, 128 (2008).

\bibitem{Cornean} M.H. Brynildsen, H.D. Cornean, \emph{Rev. Math. Phys.} \textbf{25}, 1350007 (2013).

\bibitem{PW} W. Pusz and S. L. Woronowicz, \emph{Commun. math. Phys.}
\textbf{58}, 273 (1978).

\bibitem{lieb-yngvasonPhysReport} E. H. Lieb and J. Yngvason, \emph{Phys.
Rep.} \textbf{310}, 1 (1999).

\bibitem{Emch} G.G. Emch, \emph{Algebraic Methods in Statistical Mechanics
and Quantum Field Theory} (Willey--Interscience, New York, 1972).

\bibitem{BratteliRobinson} O. Bratteli and D.W. Robinson, \emph{Operator
Algebras and Quantum Statistical Mechanics, Vol. II}, 2nd ed.
(Springer-Verlag, New York, 1996).

\bibitem{Israel} R.B. Israel, \emph{Convexity in the theory of lattice gases}
(Princeton Univ. Press, 1979).

\bibitem{Araki-Moriya} H. Araki and H. Moriya, \emph{Rev. Math. Phys.}
\textbf{15}, 93 (2003).

\bibitem{BruPedra2} J.-B. Bru and W. de Siqueira Pedra, \emph{Memoirs of the
AMS} \textbf{224}(1052) (2013).

\bibitem{BruPedra-homog} J.-B. Bru and W. de Siqueira Pedra, \emph{J. Math.
Phys.} \textbf{53}, 123301 (2012).

\bibitem{OhmI} J.-B. Bru, W. de Siqueira Pedra and C. Hertling,
\emph{Comm. Pure Appl. Math.} \textbf{68}(6) 964 (2015).

\bibitem{OhmII} J.-B. Bru, W. de Siqueira Pedra and C. Hertling,
\emph{J. Math. Phys.} \textbf{56}, 051901 (2015).

\bibitem{OhmIII} J.-B. Bru, W. de Siqueira Pedra and C. Hertling,
\emph{Archive for Rational Mechanics and
Analysis} \textbf{220}, 445 (2016).

\bibitem{OhmIV} J.-B. Bru, W. de Siqueira Pedra and C. Hertling, \emph{Rev. Math. Phys} \textbf{26}(5), 1450008 (2014).

\bibitem{OhmVbis} J.-B. Bru and W. de Siqueira Pedra, \emph{Lieb-Robinson Bounds for Multi-Commutators and Applications to Response Theory,} SpringerBriefs in Mathematical Physics (2016).

\bibitem{OhmV} J.-B. Bru and W. de Siqueira Pedra, \emph{Letters in Mathematical Physics} \textbf{106}(1), 81 (2016).

\bibitem{OhmVI} J.-B. Bru and W. de Siqueira Pedra, \emph{M3AS: Mathematical Models and Methods in Applied Sciences} \textbf{25}(14), 2587 (2015).

\bibitem{BZ} V.A. Zagrebnov and J.-B. Bru, \emph{Phys. Rep.} \textbf{350},
291 (2001).

\bibitem{spectrumexcitation1} N.M. Blagoveshchenskii et al., \emph{Phys.
Rev. B} \textbf{50}, 16550 (1994).

\bibitem{spectrumexcitation2} A. Griffin, \emph{Excitations in a
Bose-Condensated Liquid} (Cambridge Univ. Press, 1993).
\end{thebibliography}
\end{document}